\documentclass{article}
\usepackage[preprint]{colm2026_conference}

\usepackage{microtype}
\usepackage{hyperref}
\usepackage{url}
\usepackage{amsmath,amssymb}
\usepackage{graphicx}
\usepackage{booktabs}
\usepackage{multirow}
\usepackage{caption}
\usepackage{subcaption}
\usepackage{float}
\usepackage{arydshln}

\graphicspath{{figures/}}

\definecolor{darkblue}{rgb}{0, 0, 0.5}
\hypersetup{colorlinks=true, citecolor=darkblue, linkcolor=darkblue, urlcolor=darkblue}

\usepackage{lineno}

\title{Can LLMs Predict Academic Collaboration? Topology Heuristics vs.\ LLM-Based Link Prediction on Real Co-authorship Networks}

\author{%
Fan Huang \\
\mdseries Luddy School of Informatics, Computing, and Engineering \\
Indiana University Bloomington \\
\texttt{huangfan@acm.org}
\And
Munjung Kim \\
\mdseries School of Data Science \\
University of Virginia \\
\texttt{qns8tc@virginia.edu}
}

\begin{document}

\ifcolmsubmission
\linenumbers
\fi

\maketitle

\begin{abstract}
Can large language models (LLMs) predict which researchers will collaborate? We study this question through link prediction on real-world co-authorship networks from OpenAlex (9.96M authors, 108.7M edges), evaluating whether LLMs can predict future scientific collaborations using only author profiles, without access to graph structure. Using Qwen2.5-72B-Instruct across three historical eras of AI research, we find that LLMs and topology heuristics capture distinct signals and are strongest in complementary settings. On new-edge prediction under natural class imbalance, the LLM achieves AUROC 0.714--0.789, outperforming Common Neighbors, Jaccard, and Preferential Attachment, with recall up to 92.9\%; under balanced evaluation, the LLM outperforms \emph{all} topology heuristics in every era (AUROC 0.601--0.658 vs.\ best-heuristic 0.525--0.538); on continued edges, the LLM (0.687) is competitive with Adamic-Adar (0.684). Critically, 78.6--82.7\% of new collaborations occur between authors with no common neighbor---a blind spot where all topology heuristics score zero but the LLM still achieves AUROC 0.652 by reasoning from author metadata alone. A temporal metadata ablation reveals that research concepts are the dominant signal (removing concepts drops AUROC by 0.047--0.084). Providing pre-computed graph features to the LLM \emph{degrades} performance due to anchoring effects, confirming that LLMs and topology methods should operate as separate, complementary channels. A socio-cultural ablation finds that name-inferred ethnicity and institutional country do not predict collaboration beyond topology, reflecting the demographic homogeneity of AI research. A node2vec baseline achieves AUROC comparable to Adamic-Adar, establishing that LLMs access a fundamentally different information channel---author metadata---rather than encoding the same structural signal differently.
\end{abstract}

\section{Introduction}
\label{sec:introduction}

Scientific progress is shaped by the structure of collaboration networks: who works with whom determines the pace of discovery, the diffusion of ideas, and the allocation of credit~\citep{barabasi2002evolution,newman2001}. Classical link prediction methods achieve strong results on co-authorship networks by exploiting graph topology~\citep{libennowellkleinberg2007,adamic2003friends}. More recently, large language models (LLMs) have shown the ability to produce believable social behavior~\citep{park2023generative} and to reason about graph structure encoded in natural language~\citep{wang2024nlgraph,jin2024llmgraph}, prompting interest in applying LLMs to social network tasks including link prediction.

However, a critical gap remains: no study has rigorously compared LLM-based link prediction against classical topology heuristics on large-scale, real-world co-authorship networks with proper temporal evaluation. Existing evaluations often use small or synthetic datasets, lack temporal train/eval splits, or compare against weak baselines. It is therefore unclear whether LLMs can leverage author metadata (research interests, affiliations, publication records) to outperform structure-only methods. Moreover, no study has tested whether socio-cultural factors such as ethnic homophily~\citep{freeman2015,amano2016} improve LLM-based link prediction on real networks.

We address this gap with a systematic evaluation on the OpenAlex AI co-authorship network (9.96M authors, 108.7M edges) across three historical eras of AI research (2004--2023). We compare five topology heuristics, node2vec graph embeddings~\citep{grover2016node2vec}, and LLM-based prediction using Qwen2.5-72B-Instruct prompted with author profiles. We additionally test what happens when graph-derived features are provided directly to the LLM, finding that this integration degrades rather than improves prediction, revealing an anchoring effect that supports treating topology and LLM reasoning as separate channels. Specifically, we address:

\begin{enumerate}
    \item[\textbf{RQ1:}] \textbf{Topology vs.\ LLM prediction.} Which method better predicts co-authorship link formation, and how do their strengths vary across edge types (continued, new, cold-start)?

    \item[\textbf{RQ2:}] \textbf{Structural ceiling.} What fraction of new collaborations fall outside the 2-hop neighborhood that bounds all neighbor-based heuristics, and can LLMs or graph embeddings provide signal beyond this blind spot?

    \item[\textbf{RQ3:}] \textbf{Socio-cultural factors.} Does providing ethnic/cultural similarity and geographic proximity to the LLM improve link prediction beyond what topology and metadata already capture?
\end{enumerate}

\section{Related Work}
\label{sec:related}

\paragraph{Collaboration networks and link prediction.} Co-authorship networks exhibit scale-free degree distributions~\citep{barabasi2002evolution}, small-world properties~\citep{newman2001,watts1998collective}, assortative mixing~\citep{newman2002assortative}, and homophily-driven formation~\citep{mcpherson2001birds}. Generative models (SAOMs, ERGMs) capture structural regularities but not the cognitive and contextual dimensions of collaboration decisions~\citep{snijders2010,lusher2013}. \citet{libennowellkleinberg2004,libennowellkleinberg2007} formalized link prediction on co-authorship networks, establishing topology-based heuristics as strong baselines with temporal train/test evaluation. \citet{zhang2018link} proposed SEAL, using GNNs on local subgraphs. Graph embedding methods such as node2vec~\citep{grover2016node2vec} learn dense node representations from random walks. \citet{lu2011link} provide a comprehensive survey. Despite these advances, all methods operate exclusively on graph structure and cannot leverage the rich textual metadata available for scientific authors. Temporal evaluation on large-scale real co-authorship networks also remains rare.

\paragraph{LLMs for network tasks.} Recent work explores using LLMs for graph reasoning, including link prediction, by encoding structure into textual prompts~\citep{wang2024nlgraph,jin2024llmgraph}. \citet{park2023generative} demonstrated believable social behavior from LLM agents; \citet{gao2024large} surveyed LLM-empowered agent-based modeling. However, rigorous evaluation against established baselines on large-scale real networks is lacking.

\paragraph{Socio-cultural factors.} Cultural proximity strongly predicts co-authorship: \citet{freeman2015} found 2.7$\times$ same-ethnicity overrepresentation. Language barriers impose diffusion costs~\citep{amano2016,amano2023manifold,higham2025}. Institutional proximity and trust further shape collaboration~\citep{coleman1988,burt2000,lee2005impact}. However, these factors are rarely tested as inputs to link prediction models on real networks.

\section{Framework and Methodology}
\label{sec:framework}

\subsection{Data and Ground Truth Construction}

\begin{figure}[t]
    \centering
    \includegraphics[width=0.7\textwidth]{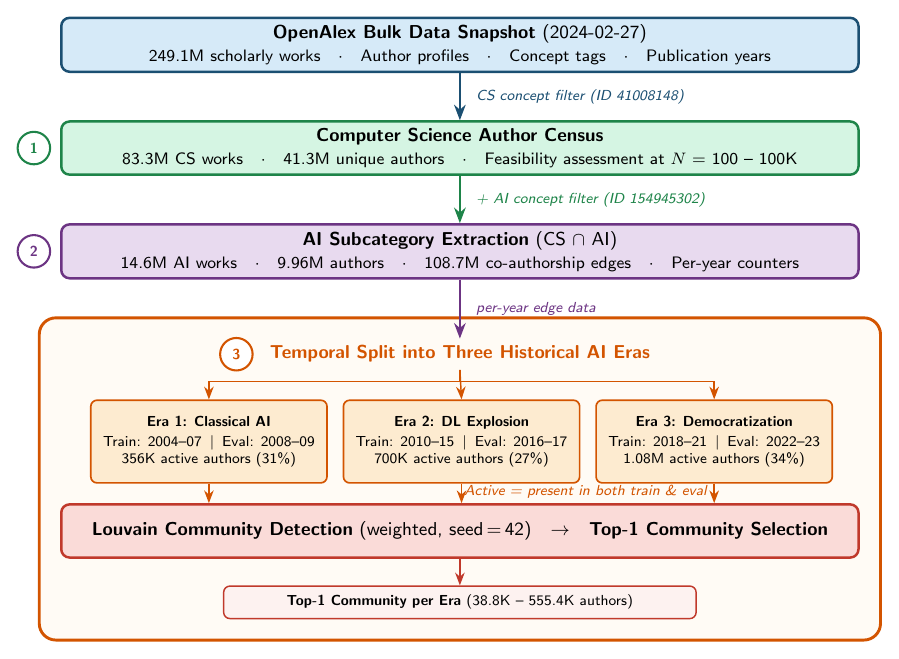}
    \caption{Ground truth construction pipeline. Three stages progressively narrow the OpenAlex corpus (249.1M works) to AI co-authorship edges, split them across three historical eras with train-eval temporal windows, detect communities via Louvain, and select the top-1 (largest) community per era for evaluation.}
    \label{fig:gt_pipeline}
\end{figure}

Co-authorship edges are extracted from the OpenAlex bulk data snapshot (2024-02-27) through a three-stage pipeline (Figure~\ref{fig:gt_pipeline}). A full scan of 249.1M works yields 14.6M AI-tagged papers, 9.96M unique authors, and 108.7M co-authorship edges with per-year temporal resolution.

The data is organized into three historical eras defined by empirical inflection points in collaboration dynamics (Table~\ref{tab:eras}; Figure~\ref{fig:era_boundaries}; see Appendix~\ref{app:data_details} for detailed statistics): \textbf{Era~1 (Classical AI, 2004--09)} with stable collaboration patterns; \textbf{Era~2 (DL Explosion, 2010--17)} with a densification regime where edges grew 2.6$\times$ while author growth remained steady; and \textbf{Era~3 (Democratization, 2018--23)} with a dilution regime where author growth outpaced edge formation. Each era uses multi-window training and a held-out evaluation window. Louvain community detection~\citep{blondel2008fast} selects the top-1 (largest) community per era (38.8K--555.4K authors). Every edge is classified into one of three types: \emph{continued} edges exist in both the training and evaluation periods (persisting collaborations); \emph{new} edges appear only in the evaluation period (novel collaborations); and \emph{dropped} edges appear only in the training period (dissolved collaborations). Among new edges, those between authors sharing at least one common neighbor are \emph{2-hop reachable}, while those with no common neighbor are \emph{cold-start}---invisible to all neighbor-based heuristics.

\begin{figure}[t]
    \centering
    \includegraphics[width=\textwidth]{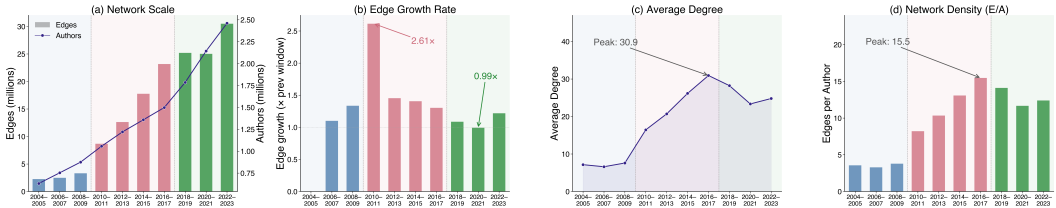}
    \caption{Empirical basis for the three-era temporal split. (a)~Network scale: edges (bars) and authors (line). (b)~Edge growth rate: the 2.61$\times$ spike marks Boundary~1; deceleration to 0.99$\times$ marks Boundary~2. (c)~Average degree peaks then declines. (d)~Edges-per-author peak then drop.}
    \label{fig:era_boundaries}
\end{figure}

\begin{table}[t]
\centering
\caption{Temporal eras and evaluation community scale. Edges classified as continued (both periods), new (eval only), or dropped (train only).}
\label{tab:eras}
\small
\begin{tabular}{llllrrr}
\toprule
\textbf{Era} & \textbf{Name} & \textbf{Train} & \textbf{Eval} & \textbf{Authors} & \textbf{New Edges} & \textbf{2-hop Cands.} \\
\midrule
1 & Classical AI     & 04--07  & 08--09 & 38,842   & 45,317   & 1.17M \\
2 & DL Explosion     & 10--15  & 16--17 & 393,843  & 914,270  & 61.6M \\
3 & Democratization  & 18--21  & 22--23 & 555,434  & 2,704,671 & 329.7M \\
\bottomrule
\end{tabular}
\end{table}

\subsection{Candidate Pair Generation}

Candidate pairs are 2-hop reachable pairs: $(u,v)$ not directly connected in training but sharing at least one common neighbor. This imposes a recall ceiling: only 17.3\% (Era~1), 20.6\% (Era~2), and 21.4\% (Era~3) of new edges occur between such pairs. The remaining 78.6--82.7\% are ``cold-start'' pairs with no shared collaborators, invisible to all neighbor-based heuristics.

\subsection{Prediction Methods}

\begin{figure}[t]
    \centering
    \includegraphics[width=0.7\textwidth]{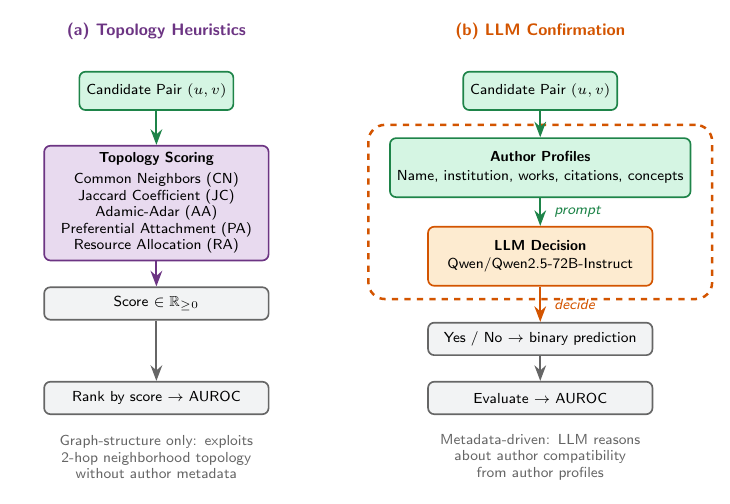}
    \caption{Two prediction methods. (a)~Topology heuristics: five structure-based scoring functions applied to candidate pairs. (b)~LLM confirmation: author profiles provided to Qwen2.5-72B for binary link prediction.}
    \label{fig:conditions}
\end{figure}

\begin{figure}[t]
    \centering
    \includegraphics[width=\textwidth]{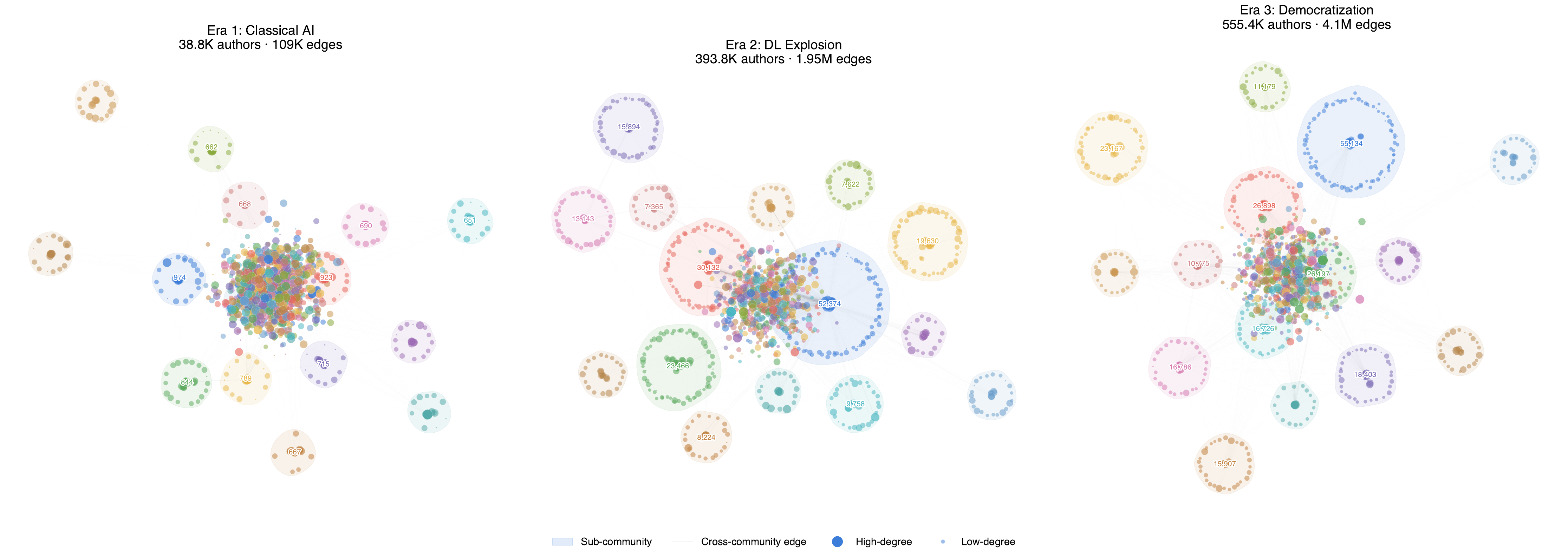}
    \caption{Experimental sample networks for each era (5,000 stratified candidate pairs). Nodes colored by sub-community; red edges indicate true future collaborations (positives). Only 47 (Era~1), 14 (Era~2), and 7 (Era~3) of 5,000 pairs are positives, illustrating the extreme class imbalance.}
    \label{fig:sample_networks}
\end{figure}

All methods operate on the same candidate pairs (Figure~\ref{fig:conditions}). Five topology heuristics score pairs from graph structure: Common Neighbors (CN), Jaccard Coefficient (JC), Adamic-Adar (AA), Preferential Attachment (PA), and Resource Allocation (RA). node2vec graph embeddings ($d{=}128$, walk length 40--80, $p{=}q{=}1$) score pairs via cosine similarity, Hadamard product, or L1/L2 distance; a sensitivity analysis over five $(p,q)$ configurations confirms robustness ($<0.01$ AUROC variation). For LLM-based prediction, each candidate pair is presented to Qwen2.5-72B-Instruct with both authors' OpenAlex profiles (name, institution, works count, cited-by count, research concepts), producing a binary yes/no prediction. We report two complementary evaluations: AA-decile stratified samples of 5,000 pairs preserving natural class imbalance (Table~\ref{tab:new_edge_imbalance}; Figure~\ref{fig:sample_networks}), and balanced samples of 250 positive and 250 negative pairs per era (Table~\ref{tab:new_edge_balanced}) for reliable per-method ranking unaffected by extreme class skew.

Author profiles use \emph{cumulative} metadata from the 2024 snapshot; research concepts---the most important field (Section~\ref{sec:temporal_ablation})---cannot be era-restricted via the OpenAlex API without aggregation bias (Appendix~\ref{app:temporal_metadata}), and a temporal ablation quantifies this design choice.

\subsection{Socio-Cultural Factor Ablation}

To address RQ3, ethnicity is inferred from author names using ethnicolr~\citep{sood2018predicting} (five categories); institutional country is resolved via ROR/Wikidata (117 countries). We evaluate standalone predictive value, LLM prompt ablation (four variants: base, +country, +ethnicity, +both), and homophily ratios (details in Appendix~\ref{app:sociocultural_details}).

\section{Results}
\label{sec:results}

\subsection{Topology vs.\ LLM Prediction (RQ1)}

\subsubsection{Topology Heuristic Performance}

Table~\ref{tab:heuristic_auroc} shows that Adamic-Adar~\citep{adamic2003friends} consistently achieves the highest AUROC across all three eras, with Resource Allocation as a close second. Preferential Attachment performs near or below chance. The stability of the AA/RA advantage across eras spanning two decades suggests that inverse-degree weighting of shared neighbors captures a durable structural signal (see Figure~\ref{fig:heuristic_auroc} in Appendix~\ref{app:additional_results}).

\begin{table}[t]
\centering
\caption{Link prediction AUROC for topology heuristics across three eras.}
\label{tab:heuristic_auroc}
\small
\begin{tabular}{lccc}
\toprule
\textbf{Heuristic} & \textbf{Era~1} & \textbf{Era~2} & \textbf{Era~3} \\
\midrule
Adamic-Adar          & \textbf{0.750} & \textbf{0.826} & \textbf{0.823} \\
Resource Allocation  & 0.735 & 0.813 & 0.803 \\
Common Neighbors     & 0.656 & 0.699 & 0.726 \\
Jaccard Coefficient  & 0.601 & 0.662 & 0.545 \\
Preferential Attach. & 0.491 & 0.437 & 0.574 \\
\bottomrule
\end{tabular}
\end{table}

node2vec achieves AUROC within 0.004--0.008 of Adamic-Adar in Eras~1--2 (0.781 vs.\ 0.773; 0.834 vs.\ 0.830; Table~\ref{tab:node2vec_auroc} in Appendix~\ref{app:additional_results}), establishing that both classical heuristics and learned embeddings reach similar performance from graph structure. This locates LLM-based prediction as accessing a distinct information channel---author metadata---rather than encoding structure differently.

\subsubsection{LLM-Based Link Prediction}

A prompt-design pilot (Appendix~\ref{app:prompt_pilot}) established that metadata-only prompts outperform prompts augmented with network statistics (AUROC 0.547 vs.\ 0.486), motivating the metadata-only design.

\paragraph{New-edge prediction.}

\begin{table}[t]
\centering
\caption{New-edge prediction AUROC under natural class imbalance (5,000 AA-decile stratified pairs per era). AA and RA dominate across all eras; the LLM outperforms CN. Era~3 contains only 7 positives, so its AUROC estimates carry high variance.}
\label{tab:new_edge_imbalance}
\small
\begin{tabular}{llrcccc}
\toprule
\textbf{Era} & \textbf{$N$} & \textbf{Pos:Neg} & \textbf{LLM} & \textbf{AA} & \textbf{RA} & \textbf{CN} \\
\midrule
Era 1 & 5,000 & 47:4,953   & 0.714 & 0.801 & \textbf{0.805} & 0.682 \\
Era 2 & 5,000 & 14:4,986   & 0.777 & 0.879 & \textbf{0.880} & 0.716 \\
Era 3 & 5,000 & 7:4,993    & 0.789 & \textbf{0.879} & 0.878 & 0.758 \\
\bottomrule
\end{tabular}
\end{table}

\begin{table}[t]
\centering
\caption{New-edge prediction AUROC under balanced sampling (250 positive, 250 negative, AA-quintile stratified). The LLM outperforms all topology heuristics in every era.}
\label{tab:new_edge_balanced}
\small
\begin{tabular}{llrcccc}
\toprule
\textbf{Era} & \textbf{$N$} & \textbf{Pos:Neg} & \textbf{LLM} & \textbf{AA} & \textbf{RA} & \textbf{CN} \\
\midrule
Era 1 & 500 & 250:250 & \textbf{0.658} & 0.533 & 0.531 & 0.538 \\
Era 2 & 500 & 250:250 & \textbf{0.615} & 0.520 & 0.524 & 0.525 \\
Era 3 & 500 & 250:250 & \textbf{0.601} & 0.509 & 0.512 & 0.533 \\
\bottomrule
\end{tabular}
\end{table}

Three findings stand out (Tables~\ref{tab:new_edge_imbalance}--\ref{tab:new_edge_balanced}). First, under natural imbalance (Table~\ref{tab:new_edge_imbalance}), AA and RA dominate across all eras, while the LLM consistently outperforms CN (e.g., 0.714 vs.\ 0.682 in Era~1; 0.777 vs.\ 0.716 in Era~2). Second, LLM recall on new edges is 91.5\% in Era~1 (43/47) and 92.9\% in Era~2 (13/14), indicating strong screening capability. Third, under balanced evaluation (Table~\ref{tab:new_edge_balanced}), the LLM \textbf{outperforms all topology heuristics in every era} (0.658 vs.\ best-heuristic 0.538 in Era~1; 0.615 vs.\ 0.525 in Era~2; 0.601 vs.\ 0.533 in Era~3), suggesting the LLM captures signals beyond structural proximity that are masked by class imbalance. However, the LLM exhibits uniformly high false positive rates (0.6--0.9) across all AA deciles, indicating poor calibration (Figure~\ref{fig:calibration_decile} in Appendix~\ref{app:additional_results}).

\paragraph{Continued-edge prediction.} For predicting persistence of existing collaborations (Era~1, $N{=}150$), the LLM (0.687) is essentially tied with Adamic-Adar (0.684) and only marginally below Resource Allocation (0.706), demonstrating that LLM reasoning captures signals relevant to tie persistence comparable to topology methods (Table~\ref{tab:continued_edge} in Appendix~\ref{app:additional_results}).

\subsubsection{Temporal Metadata Ablation}
\label{sec:temporal_ablation}

To quantify whether cumulative metadata introduces temporal leakage, we ran a controlled ablation (Table~\ref{tab:temporal_ablation}).

\begin{table}[t]
\centering
\caption{Temporal metadata ablation AUROC (50,000 pairs per era). $\Delta$(conc.) = effect of removing concepts; $\Delta$(era) = effect of era-restricting counts and institution.}
\label{tab:temporal_ablation}
\small
\begin{tabular}{lccccc}
\toprule
\textbf{Era} & \textbf{w/ Conc.} & \textbf{No-Conc.} & \textbf{Era-Restr.} & \textbf{$\Delta$(conc.)} & \textbf{$\Delta$(era)} \\
\midrule
Era~1 & \textbf{0.573} & 0.526 & 0.544 & $+$0.047 & $+$0.018 \\
Era~2 & \textbf{0.694} & 0.610 & 0.562 & $+$0.084 & $-$0.048 \\
Era~3 & \textbf{0.707} & 0.627 & 0.570 & $+$0.080 & $-$0.057 \\
\bottomrule
\end{tabular}
\end{table}

\textbf{Concepts are the dominant metadata signal}: removing them drops AUROC by 0.047--0.084, the largest single-factor effect. Era-restricting other fields produces smaller, mixed effects ($-$0.057 to $+$0.018). The concept effect dominates the leakage effect by 1.4--2.6$\times$, validating the cumulative metadata approach since concepts cannot be era-restricted without aggregation bias (Appendix~\ref{app:temporal_metadata}).

\subsubsection{Method Agreement and Complementarity}

To understand \emph{where} each method succeeds and fails, we classify all 68 positive pairs by whether AA and the LLM correctly predict them (using per-era median-positive AA threshold and LLM threshold 0.5). Figure~\ref{fig:agreement_scatter} visualizes the four agreement categories; detailed case studies are in Appendix~\ref{app:examples}.

\begin{figure}[t]
    \centering
    \includegraphics[width=0.65\textwidth]{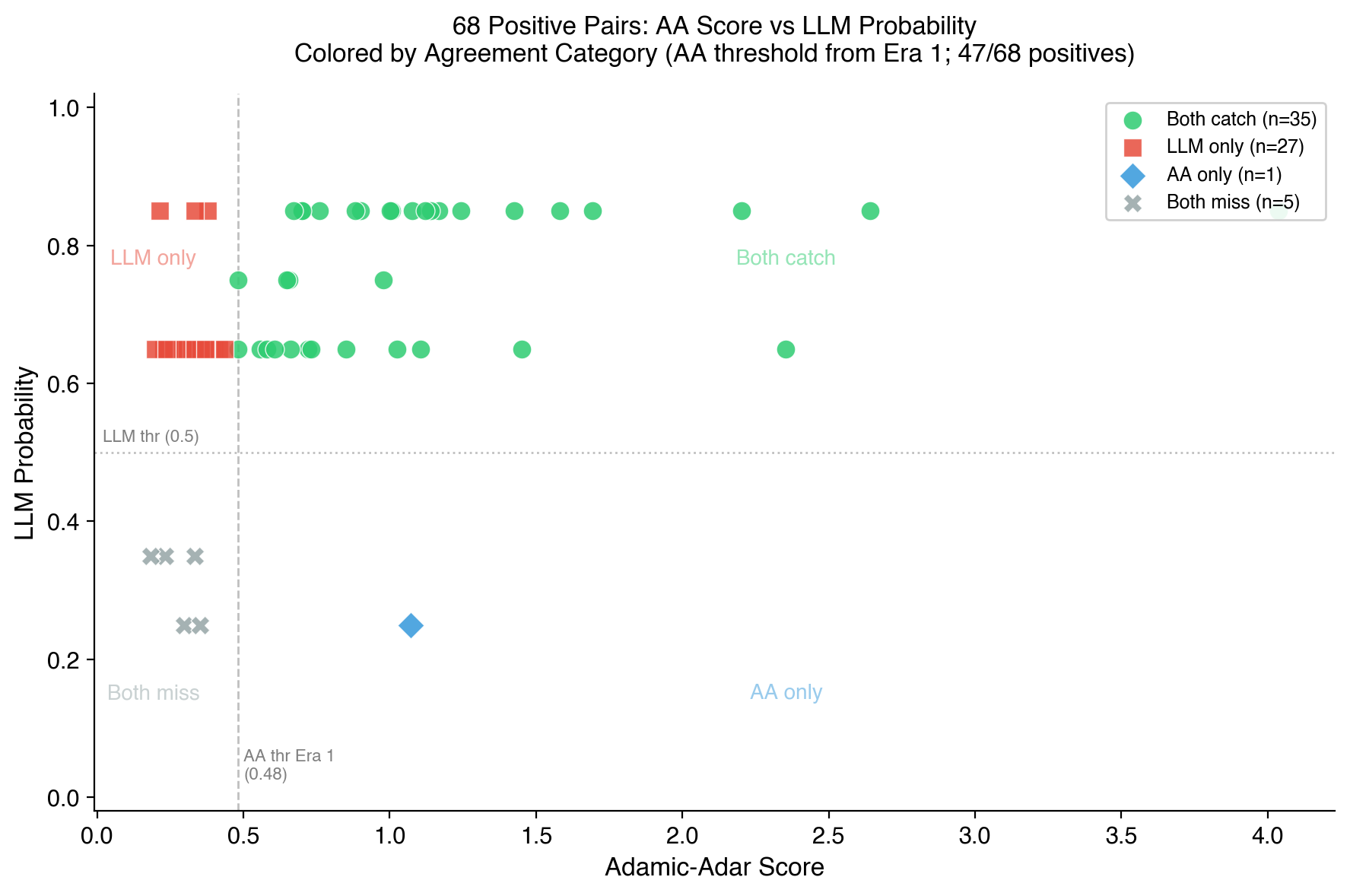}
    \caption{68 positive pairs plotted by AA score vs.\ LLM probability, colored by agreement category. The LLM catches 40\% of positives that AA misses (red squares, lower-left quadrant), while only 1 pair is caught by AA alone. Dashed/dotted lines show Era~1 AA threshold (0.48) and LLM threshold (0.5).}
    \label{fig:agreement_scatter}
\end{figure}

Four patterns emerge: (1)~\textbf{Both catch} (35/68, 51\%): pairs with strong structural \emph{and} metadata signals (median AA\,=\,0.98, concept overlap\,=\,0.43, 83\% intra-community). (2)~\textbf{LLM only} (27/68, 40\%): pairs with low AA (median 0.32, CN\,=\,1) but the LLM detects shared research focus; 63\% are cross-community, directly validating the complementarity thesis. (3)~\textbf{AA only} (1/68): a single pair with zero concept overlap where the LLM cannot compensate for missing metadata. (4)~\textbf{Both miss} (5/68, 7\%): zero concept overlap, cross-community, representing an irreducible prediction floor of serendipitous collaborations.

The per-edge-type results reveal that aggregate AUROC obscures meaningful variation. Three key patterns emerge. First, \emph{anchoring on numerical features}: including pre-computed network statistics in LLM prompts degraded performance from 0.547 to 0.486 (Appendix~\ref{app:prompt_pilot}); LLMs anchored on salient AA scores (Spearman $\rho = 0.42$) rather than reasoning independently. Second, \emph{hard-negative sampling}: the general evaluation draws negatives from top AA deciles, creating an artificially difficult task; per-edge-type analysis reveals substantially better LLM performance on natural sub-tasks. Third, \emph{complementary signal}: LLM probabilities correlate positively with AA scores ($\rho = 0.25$--$0.32$), suggesting overlapping but not identical signal. An ensemble exploiting both channels could leverage their complementary strengths.

\subsection{Structural Ceiling and Cold-Start Prediction (RQ2)}

Only 17.3--21.4\% of new collaborations occur between authors sharing a common neighbor. The remaining 78.6--82.7\% involve cold-start pairs where \emph{all} neighbor-based heuristics score identically zero~\citep{libennowellkleinberg2007}. Even a perfect neighbor-based heuristic can recover at most one-fifth of new collaborations. Table~\ref{tab:cold_start} evaluates multiple method classes on cold-start pairs.

\begin{table}[t]
\centering
\caption{Cold-start prediction AUROC (top-1,000 authors, 500 sampled pairs, 1 positive per era). CN-based heuristics score exactly 0.500. Era~3 is omitted because zero cold-start positives exist among the top-1,000 hubs (cold-start collaborations in Era~3 occur only among lower-degree authors).}
\label{tab:cold_start}
\small
\begin{tabular}{llcc}
\toprule
\textbf{Category} & \textbf{Method} & \textbf{Era~1} & \textbf{Era~2} \\
\midrule
\multirow{2}{*}{Topology}
    & AA / CN / JC / RA & 0.500 & 0.500 \\
    & Pref.\ Attachment & 0.095 & 0.848 \\
\midrule
\multirow{2}{*}{Embedding}
    & N2V-Hadamard      & 0.651 & \textbf{0.916} \\
    & N2V-Cosine        & 0.154 & 0.848 \\
\midrule
\multirow{3}{*}{Metadata}
    & $\text{cited\_by}(u) \times \text{cited\_by}(v)$  & \textbf{0.876} & \textbf{0.942} \\
    & $|\text{concepts}(u) \cap \text{concepts}(v)|$   & 0.779 & 0.801 \\
    & $\text{works}(u) \times \text{works}(v)$     & 0.754 & 0.934 \\
\midrule
LLM & Qwen-72B          & 0.652 & 0.869 \\
\bottomrule
\end{tabular}
\end{table}

\begin{figure}[t]
    \centering
    \includegraphics[width=0.8\textwidth]{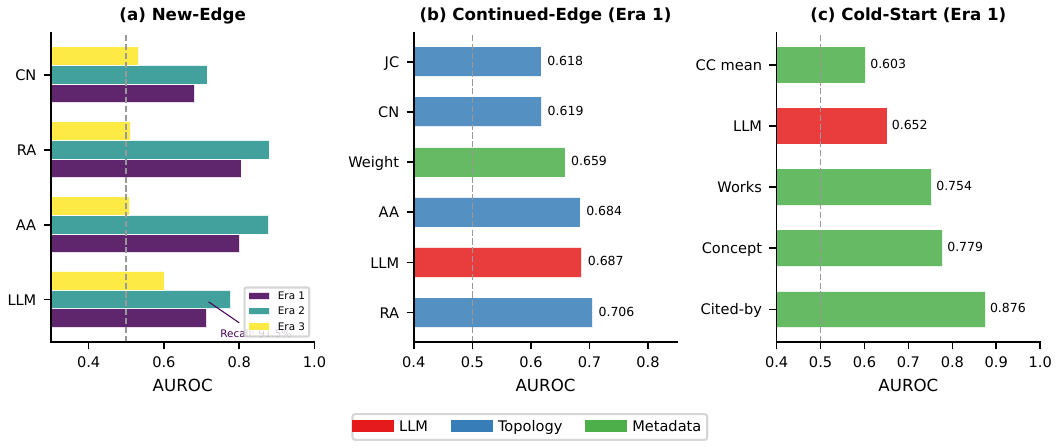}
    \caption{AUROC by edge type and prediction method (natural imbalance). (a)~New edges: AA and RA lead, but LLM outperforms CN and JC; under balanced sampling the LLM outperforms all heuristics (Table~\ref{tab:new_edge_balanced}). (b)~Continued edges: LLM (0.687) is competitive with AA (0.684). (c)~Cold-start: metadata features dominate; LLM (0.652) provides signal where topology scores zero. Dashed line = random.}
    \label{fig:edge_type_performance}
\end{figure}

We additionally evaluate three metadata-derived baselines (product of citation counts, shared research concepts, product of publication counts) to test how much signal resides in raw author metadata alone. CN-based heuristics confirm 0.500 (random) as expected; node2vec provides meaningful signal through Hadamard (0.651--0.916) but is operator-dependent; metadata baselines provide the strongest cold-start signal overall (0.754--0.942), confirming that citation prominence and topic overlap are the primary basis for predicting collaborations between structurally disconnected researchers. The LLM achieves consistent performance (0.652--0.869) competitive with the best embedding operator, capturing similar signals through natural language reasoning over author profiles. Cold-start is a long-tail phenomenon (Appendix~\ref{app:cold_start_details}): characterized by low author degree, high structural distance (median 4--5 hops), and cross-community bridging (81--97\% vs.\ 39--67\% for 2-hop edges)---precisely the population for whom topology methods have the least signal. Figure~\ref{fig:edge_type_performance} consolidates the per-edge-type results: no single method dominates, with topology leading on 2-hop ranking, the LLM competitive on continued edges, and metadata features strongest for cold-start pairs.

\subsection{Socio-Cultural Factor Analysis (RQ3)}

Ethnicity predictions (79.7\% Asian, 17.0\% White) and institutional country assignments (117 countries) were obtained for all 20,389 authors in Era~1 (100\% coverage). All three socio-cultural features achieve AUROC \emph{below} 0.5 as standalone predictors (ethnicity: 0.475, country: 0.448, continent: 0.445), indicating slight anti-homophily within the 2-hop candidate pool (Table~\ref{tab:sociocultural_auroc} in Appendix~\ref{app:sociocultural_details}).

\begin{table}[t]
\centering
\caption{LLM prompt ablation with socio-cultural factors (Era~1, $N{=}5{,}000$, 47 positives).}
\label{tab:ablation_auroc}
\small
\begin{tabular}{lcc}
\toprule
\textbf{Prompt Variant} & \textbf{AUROC} & \textbf{$\Delta$ vs.\ Base} \\
\midrule
Base (metadata only)       & \textbf{0.699} & --- \\
+ Country                  & 0.686 & $-0.013$ \\
+ Ethnicity                & 0.686 & $-0.013$ \\
+ Both                     & 0.683 & $-0.016$ \\
\bottomrule
\end{tabular}
\end{table}

Adding socio-cultural context consistently \emph{degrades} AUROC by 1.3--1.6 points (Table~\ref{tab:ablation_auroc}). Homophily ratios fall below 1.0 across all features and eras (Appendix~\ref{app:sociocultural_details}), indicating that within the structurally proximate 2-hop pool, collaborators \emph{bridge across} groups rather than cluster within them. This contrasts with population-level findings~\citep{freeman2015} and reflects both the candidate selection conditioning on structural proximity and the demographic homogeneity of AI research.

\section{Discussion}
\label{sec:discussion}

Adamic-Adar achieves AUROC 0.750--0.826 with no metadata and minimal computation; node2vec's comparable performance (0.781--0.834) confirms that structural signal is well-captured by multiple approaches. LLMs thus add not a better encoding of structure, but access to an entirely different information source. Topology dominates within 2-hop neighborhoods under natural imbalance, yet this covers only 17--21\% of new edges. Under balanced evaluation, the LLM outperforms all heuristics in every era (0.601--0.658 vs.\ best-heuristic 0.525--0.538), suggesting topology's advantage is partly an artifact of class skew. The LLM also provides strong new-edge recall (91.5--92.9\%), competitive continued-edge prediction (0.687 vs.\ AA 0.684), and consistent cold-start signal (0.652--0.869), motivating an ensemble that pairs AA for 2-hop ranking with LLM screening for the 78.6--82.7\% of new edges invisible to neighbor-based methods.

Critically, providing graph features to the LLM \emph{degrades} performance (0.547$\to$0.486), as LLMs anchor on salient numerical scores (Spearman $\rho{=}0.42$) rather than reasoning independently; topology and LLMs should therefore operate as \emph{separate parallel channels}, not fused into a single input. Research concepts are the dominant metadata signal: removing them drops AUROC by 0.047--0.084, exceeding the temporal leakage effect by 1.4--2.6$\times$, consistent with homophily theory~\citep{mcpherson2001birds}. Within structurally proximate candidate pools in a demographically homogeneous field (79.7\% Asian), socio-cultural features carry no additional predictive signal, reflecting candidate selection effects, demographic concentration, and information redundancy (institution names already encode geography), bounding rather than contradicting population-level homophily findings~\citep{freeman2015}.

\section{Conclusion and Future Work}
\label{sec:conclusion}

On the OpenAlex AI co-authorship network (9.96M authors, 108.7M edges, three eras), we find that topology heuristics and LLMs capture complementary signals: (1)~structure-based methods achieve strong 2-hop ranking (AA AUROC 0.750--0.826, node2vec 0.781--0.834); (2)~LLMs outperform all heuristics under balanced evaluation (0.601--0.658 vs.\ 0.525--0.538), provide strong new-edge recall (91.5--92.9\%) and consistent cold-start signal (0.652--0.869), but degrade when given graph features; (3)~78.6--82.7\% of new collaborations are cold-start; (4)~research concepts are the dominant metadata signal ($\Delta$AUROC 0.047--0.084); (5)~socio-cultural factors do not improve prediction. LLMs are strong at screening novel and cold-start collaborations from metadata, but not a replacement for structure-based methods within established neighborhoods.

Several directions for future work follow naturally. The complementary strengths identified here motivate a \emph{two-stage ensemble} combining Adamic-Adar for 2-hop ranking with LLM screening for cold-start pairs, potentially addressing the 78.6--82.7\% of new edges invisible to neighbor-based methods. The LLM's poor calibration (high false positive rates across all AA deciles) suggests that fine-tuned smaller models or confidence-calibrated prompting strategies could improve precision at lower inference cost. Incorporating richer metadata may strengthen the metadata channel further. Testing finer-grained ethnicity taxonomies in more demographically diverse fields (e.g., biomedical research, social sciences) would establish whether the socio-cultural null result generalizes or is specific to the demographic homogeneity of AI research. Finally, extending the framework to dynamic link prediction with continuous-time models could capture the temporal dynamics of collaboration formation more precisely.

\section*{Acknowledgments}
We gratefully acknowledge the open-source resources that made this work possible: OpenAlex for providing open bibliographic data and API access, the Qwen team for releasing Qwen2.5-72B-Instruct as an open-weight model, and the developers of node2vec, ethnicolr, and the broader open-source scientific Python ecosystem. AI writing assistants (Claude, ChatGPT) were used solely for grammar checking and proofreading.

\section*{Ethics Statement}
This study uses publicly available bibliographic data from OpenAlex and does not involve human subjects or private information. Ethnicity predictions are inferred from author names using ethnicolr for analytical purposes only, to test whether demographic homophily improves link prediction; we acknowledge the inherent limitations and potential for misclassification in name-based ethnicity inference. Our findings show that such features do not improve prediction in this setting, reinforcing that collaboration prediction systems should not rely on demographic attributes. All data processing follows OpenAlex's open data license and intended use.

\section*{Reproducibility Statement}
The framework, including all analysis pipelines and extracted datasets, will be released as open-source upon publication.

\bibliography{references}

@article{newman2001,
  title={The Structure of Scientific Collaboration Networks},
  author={Newman, Mark E. J.},
  journal={Proceedings of the National Academy of Sciences},
  volume={98},
  number={2},
  pages={404--409},
  year={2001},
  publisher={National Academy of Sciences}
}

@article{newman2002assortative,
  title={Assortative Mixing in Networks},
  author={Newman, Mark E. J.},
  journal={Physical Review Letters},
  volume={89},
  number={20},
  pages={208701},
  year={2002},
  publisher={American Physical Society}
}

@article{burt2000,
  title={The Network Structure of Social Capital},
  author={Burt, Ronald S.},
  journal={Research in Organizational Behavior},
  volume={22},
  pages={345--423},
  year={2000},
  publisher={Elsevier}
}

@article{lee2005impact,
  title={The Impact of Research Collaboration on Scientific Productivity},
  author={Lee, Sooho and Bozeman, Barry},
  journal={Social Studies of Science},
  volume={35},
  number={5},
  pages={673--702},
  year={2005},
  publisher={SAGE}
}

@inproceedings{park2023generative,
  title={Generative Agents: Interactive Simulacra of Human Behavior},
  author={Park, Joon Sung and O'Brien, Joseph C and Cai, Carrie J and Morris, Meredith Ringel and Liang, Percy and Bernstein, Michael S},
  booktitle={Proceedings of the 36th Annual ACM Symposium on User Interface Software and Technology (UIST '23)},
  pages={1--22},
  year={2023},
  publisher={ACM}
}

@article{gao2024large,
  title={Large Language Models Empowered Agent-based Modeling and Simulation: A Survey and Perspectives},
  author={Gao, Chen and Lan, Xiaochong and Li, Nian and Yuan, Yuan and Ding, Jingtao and Zhou, Zhilun and others},
  journal={Humanities and Social Sciences Communications},
  volume={11},
  pages={1--24},
  year={2024},
  publisher={Nature Publishing Group},
  doi={10.1057/s41599-024-03611-3}
}

@article{barabasi2002evolution,
  title={Evolution of the Social Network of Scientific Collaborations},
  author={Barab{\'a}si, Albert-L{\'a}szl{\'o} and Jeong, Hawoong and N{\'e}da, Zolt{\'a}n and Ravasz, Erzs{\'e}bet and Schubert, Andr{\'a}s and Vicsek, Tam{\'a}s},
  journal={Physica A: Statistical Mechanics and its Applications},
  volume={311},
  number={3-4},
  pages={590--614},
  year={2002},
  publisher={Elsevier}
}

@article{mcpherson2001birds,
  title={Birds of a Feather: Homophily in Social Networks},
  author={McPherson, Miller and Smith-Lovin, Lynn and Cook, James M},
  journal={Annual Review of Sociology},
  volume={27},
  number={1},
  pages={415--444},
  year={2001},
  publisher={Annual Reviews}
}

@article{snijders2010,
  title={Introduction to Stochastic Actor-Based Models for Network Dynamics},
  author={Snijders, Tom AB and Van de Bunt, Gerhard G and Steglich, Christian EG},
  journal={Social Networks},
  volume={32},
  number={1},
  pages={44--60},
  year={2010},
  publisher={Elsevier}
}

@book{lusher2013,
  title={Exponential Random Graph Models for Social Networks: Theory, Methods, and Applications},
  author={Lusher, Dean and Koskinen, Johan and Robins, Garry},
  year={2013},
  publisher={Cambridge University Press}
}

@article{amano2023manifold,
  title={The Manifold Costs of Being a Non-Native English Speaker in Science},
  author={Amano, Tatsuya and Rios Rojas, Celia and Boum II, Yap and others},
  journal={PLoS Biology},
  volume={21},
  number={7},
  pages={e3002184},
  year={2023},
  publisher={Public Library of Science}
}

@article{watts1998collective,
  title={Collective Dynamics of `Small-World' Networks},
  author={Watts, Duncan J and Strogatz, Steven H},
  journal={Nature},
  volume={393},
  number={6684},
  pages={440--442},
  year={1998},
  publisher={Nature Publishing Group}
}

@article{amano2016,
  title={Languages Are Still a Major Barrier to Global Science},
  author={Amano, Tatsuya and Gonz{\'a}lez-Varo, Juan P. and Sutherland, William J.},
  journal={PLoS Biology},
  volume={14},
  number={12},
  pages={e2000933},
  year={2016},
  publisher={Public Library of Science}
}

@article{coleman1988,
  title={Social Capital in the Creation of Human Capital},
  author={Coleman, James S.},
  journal={American Journal of Sociology},
  volume={94},
  pages={S95--S120},
  year={1988},
  publisher={University of Chicago Press}
}

@article{freeman2015,
  title={Collaborating with People Like Me: Ethnic Coauthorship within the United States},
  author={Freeman, Richard B. and Huang, Wei},
  journal={Journal of Labor Economics},
  volume={33},
  number={S1},
  pages={S289--S318},
  year={2015},
  publisher={University of Chicago Press}
}

@article{higham2025,
  title={Language Barriers and the Speed of International Knowledge Diffusion},
  author={Higham, Kyle and Nagaoka, Sadao},
  journal={Nature Human Behaviour},
  year={2026},
  publisher={Nature Publishing Group},
  doi={10.1038/s41562-025-02367-3}
}

@article{blondel2008fast,
  title={Fast Unfolding of Communities in Large Networks},
  author={Blondel, Vincent D and Guillaume, Jean-Loup and Lambiotte, Renaud and Lefebvre, Etienne},
  journal={Journal of Statistical Mechanics: Theory and Experiment},
  volume={2008},
  number={10},
  pages={P10008},
  year={2008},
  publisher={IOP Publishing}
}

@article{adamic2003friends,
  title={Friends and Neighbors on the Web},
  author={Adamic, Lada A. and Adar, Eytan},
  journal={Social Networks},
  volume={25},
  number={3},
  pages={211--230},
  year={2003},
  publisher={Elsevier}
}

@inproceedings{wang2024nlgraph,
  title={Can Language Models Solve Graph Problems in Natural Language?},
  author={Wang, Heng and Feng, Shangbin and He, Tianxing and Tan, Zhaoxuan and Han, Xiaochuang and Tsvetkov, Yulia},
  booktitle={Advances in Neural Information Processing Systems},
  volume={36},
  year={2023}
}

@article{jin2024llmgraph,
  title={Large Language Models on Graphs: A Comprehensive Survey},
  author={Jin, Bowen and Liu, Gang and Han, Chi and Jiang, Meng and Ji, Heng and Han, Jiawei},
  journal={IEEE Transactions on Knowledge and Data Engineering},
  volume={36},
  number={12},
  pages={8622--8642},
  year={2024},
  publisher={IEEE},
  doi={10.1109/TKDE.2024.3469578}
}

@article{sood2018predicting,
  title={Predicting Race and Ethnicity From the Sequence of Characters in a Name},
  author={Sood, Gaurav and Laohaprapanon, Suriyan},
  journal={arXiv preprint arXiv:1805.02109},
  year={2018}
}

@article{libennowellkleinberg2007,
  title={The Link-Prediction Problem for Social Networks},
  author={Liben-Nowell, David and Kleinberg, Jon},
  journal={Journal of the American Society for Information Science and Technology},
  volume={58},
  number={7},
  pages={1019--1031},
  year={2007},
  publisher={Wiley}
}

@inproceedings{zhang2018link,
  title={Link Prediction Based on Graph Neural Networks},
  author={Zhang, Muhan and Chen, Yixin},
  booktitle={Advances in Neural Information Processing Systems},
  volume={31},
  pages={5165--5175},
  year={2018}
}

@inproceedings{libennowellkleinberg2004,
  title={The Link Prediction Problem for Social Networks},
  author={Liben-Nowell, David and Kleinberg, Jon},
  booktitle={Proceedings of the Twelfth International Conference on Information and Knowledge Management (CIKM '03)},
  pages={556--559},
  year={2003},
  publisher={ACM}
}

@inproceedings{grover2016node2vec,
  title={node2vec: Scalable Feature Learning for Networks},
  author={Grover, Aditya and Leskovec, Jure},
  booktitle={Proceedings of the 22nd ACM SIGKDD International Conference on Knowledge Discovery and Data Mining},
  pages={855--864},
  year={2016},
  organization={ACM}
}

@inproceedings{krizhevsky2012imagenet,
  title={ImageNet Classification with Deep Convolutional Neural Networks},
  author={Krizhevsky, Alex and Sutskever, Ilya and Hinton, Geoffrey E.},
  booktitle={Advances in Neural Information Processing Systems},
  volume={25},
  pages={1097--1105},
  year={2012}
}

@article{lecun2015deep,
  title={Deep Learning},
  author={LeCun, Yann and Bengio, Yoshua and Hinton, Geoffrey},
  journal={Nature},
  volume={521},
  number={7553},
  pages={436--444},
  year={2015},
  publisher={Nature Publishing Group}
}

@article{frank2019evolution,
  title={The Evolution of Citation Graphs in Artificial Intelligence Research},
  author={Frank, Morgan R. and Wang, Dashun and Cebrian, Manuel and Rahwan, Iyad},
  journal={Nature Machine Intelligence},
  volume={1},
  number={2},
  pages={79--85},
  year={2019},
  publisher={Nature Publishing Group}
}

@article{lu2011link,
  title={Link Prediction in Complex Networks: A Survey},
  author={L{\"u}, Linyuan and Zhou, Tao},
  journal={Physica A: Statistical Mechanics and Its Applications},
  volume={390},
  number={6},
  pages={1150--1170},
  year={2011},
  publisher={Elsevier}
}

@incollection{glanzel2004coauthorship,
  title={Analysing Scientific Networks Through Co-authorship},
  author={Gl{\"a}nzel, Wolfgang and Schubert, Andr{\'a}s},
  booktitle={Handbook of Quantitative Science and Technology Research},
  pages={257--276},
  year={2004},
  publisher={Springer}
}

@article{visser2021comparison,
  title={Large-scale Comparison of Bibliographic Data Sources: {Scopus}, {Web of Science}, {Dimensions}, {Crossref}, and {Microsoft Academic}},
  author={Visser, Martijn and van Eck, Nees Jan and Waltman, Ludo},
  journal={Quantitative Science Studies},
  volume={2},
  number={1},
  pages={20--41},
  year={2021},
  publisher={MIT Press}
}
\bibliographystyle{colm2026_conference}

\clearpage

\appendix

\section{Data and Era Construction Details}
\label{app:data_details}

\paragraph{Three-stage pipeline.} Stage~1 (Domain Census) scans all 249.1M OpenAlex works to identify 83.3M CS-tagged works by 41.3M researchers. Stage~2 (Subcategory Extraction) filters for works tagged with both CS (level-0) and AI (level-1), yielding 14.6M AI works, 9.96M unique authors, and 108.7M co-authorship edges. Stage~3 (Temporal Split and Community Selection) organizes per-year edge data into three eras.

\paragraph{Two-year window justification.} We adopt two-year aggregation windows, supported by: (1)~established link prediction practice~\citep{libennowellkleinberg2007}; (2)~community consensus that 2--5 year windows are standard~\citep{lu2011link}; (3)~publication cycle lags of 6--18 months~\citep{glanzel2004coauthorship}; (4)~database date-assignment artifacts~\citep{visser2021comparison}. Our data confirm: single-year edge counts exhibit 0.34$\times$ to 3.73$\times$ fluctuations.

\paragraph{Era boundaries.} Boundary~1 (2009$\to$2010): edges jump 2.6$\times$ while author growth stays at 1.21$\times$, coinciding with the deep learning paradigm shift~\citep{krizhevsky2012imagenet,lecun2015deep}. Boundary~2 (2017$\to$2018): edge growth decelerates to 0.99--1.09$\times$ while author growth accelerates to 1.19--1.20$\times$, consistent with mass participation~\citep{frank2019evolution}.

\begin{table}[H]
\centering
\caption{OpenAlex AI co-authorship network statistics by two-year window. Dashed lines mark era boundaries.}
\label{tab:year_stats}
\small
\begin{tabular}{lrrrrr}
\toprule
\textbf{Window} & \textbf{Authors} & \textbf{Edges} & \textbf{Avg Deg} & \textbf{Edge Gr.} & \textbf{E/A} \\
\midrule
2004--05 & 631{,}767  & 2{,}257{,}606  & 7.15  & ---                    & 3.6 \\
2006--07 & 753{,}373  & 2{,}486{,}310  & 6.60  & 1.10$\times$           & 3.3 \\
2008--09 & 875{,}530  & 3{,}322{,}828  & 7.59  & 1.34$\times$           & 3.8 \\
\hdashline
2010--11 & 1{,}056{,}752 & 8{,}686{,}468  & 16.44 & \textbf{2.61$\times$} & 8.2 \\
2012--13 & 1{,}220{,}422 & 12{,}628{,}718 & 20.70 & 1.45$\times$          & 10.3 \\
2014--15 & 1{,}358{,}721 & 17{,}771{,}014 & 26.16 & 1.41$\times$          & 13.1 \\
2016--17 & 1{,}498{,}564 & 23{,}175{,}943 & 30.93 & 1.30$\times$          & 15.5 \\
\hdashline
2018--19 & 1{,}785{,}519 & 25{,}205{,}216 & 28.23 & 1.09$\times$          & 14.1 \\
2020--21 & 2{,}143{,}510 & 25{,}029{,}315 & 23.35 & 0.99$\times$          & 11.7 \\
2022--23 & 2{,}460{,}447 & 30{,}508{,}876 & 24.80 & 1.22$\times$          & 12.4 \\
\bottomrule
\multicolumn{6}{l}{\footnotesize E/A = edges per author (density proxy). Edge Gr.\ = ratio to previous window.}
\end{tabular}
\end{table}

\paragraph{Baseline methods.} Four baselines establish performance bounds: Random (lower bound), Persistent (training edges ranked by weight), Common Neighbors ($|N(u) \cap N(v)|$), and Preferential Attachment ($\deg(u) \times \deg(v)$). The persistence signal is strong: continued edges comprise 23--29\% of training edges, while 59--74\% of evaluation edges are new.

\section{Experimental Setup Details}
\label{app:experimental_details}

\begin{figure}[H]
    \centering
    \includegraphics[width=\textwidth]{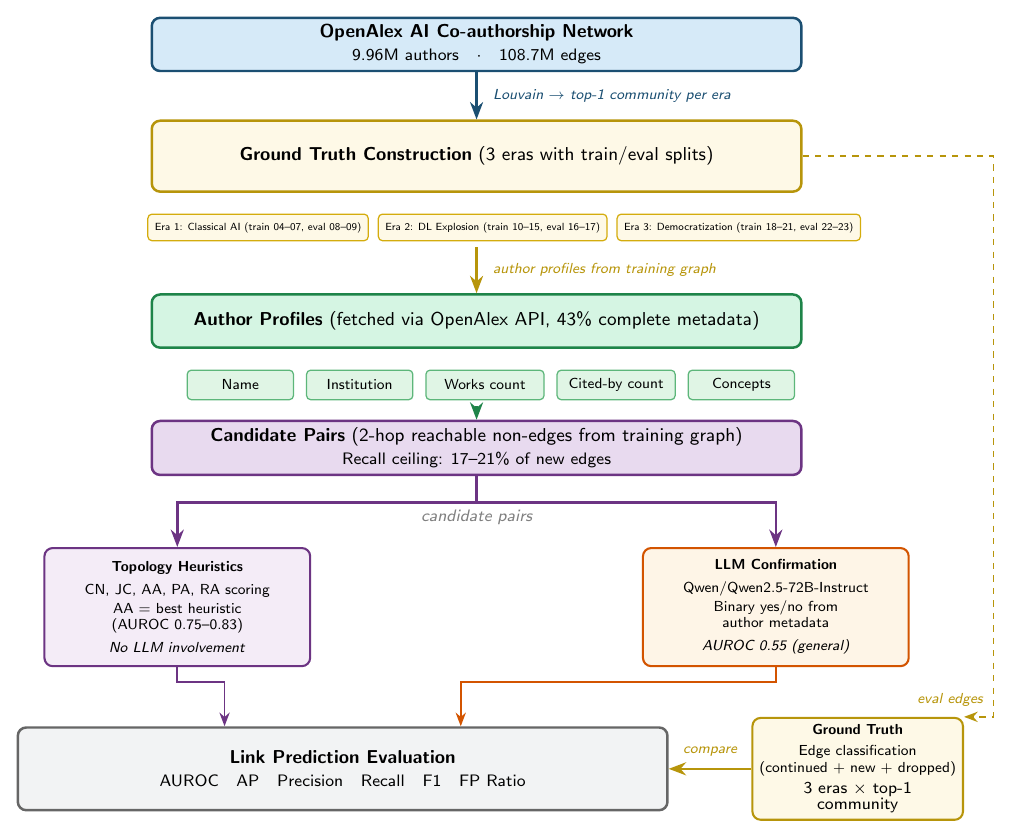}
    \caption{Link prediction evaluation pipeline. The OpenAlex AI co-authorship network is split into three historical eras with train/eval periods. Author profiles are fetched via the OpenAlex API. Candidate pairs are scored by topology heuristics or classified by an LLM.}
    \label{fig:framework}
\end{figure}

\begin{figure}[H]
    \centering
    \includegraphics[width=\textwidth]{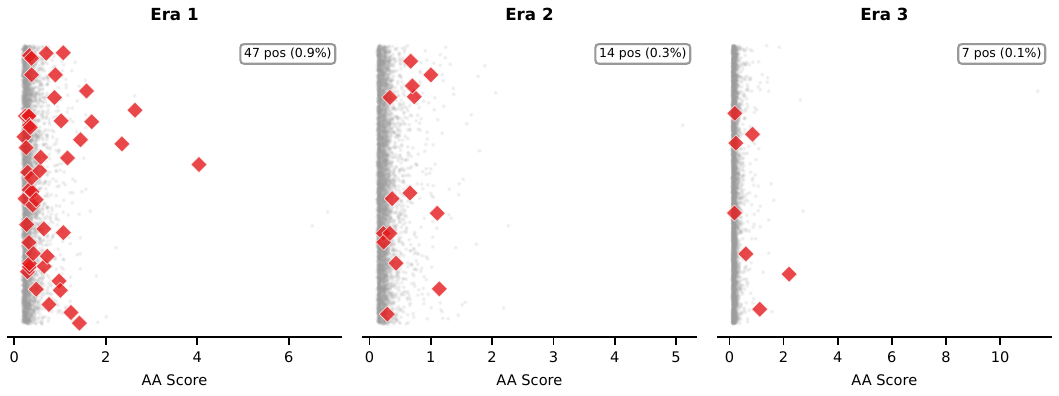}
    \caption{Distribution of Adamic-Adar scores for 5,000 stratified candidate pairs per era. Positives cluster at higher AA scores but remain needle-in-haystack rare: 47 (Era~1), 14 (Era~2), and 7 (Era~3) out of 5,000 pairs.}
    \label{fig:sample_aa_strip}
\end{figure}

\paragraph{Computational considerations.} The largest community (555.4K authors) contains 329.7M 2-hop candidate pairs; at 22.5 requests/minute, exhaustive LLM evaluation would require 976,000 GPU-hours. We therefore evaluate on sampled pairs and report AUROC, which is robust to sampling.

\section{Prompt Design Pilot}
\label{app:prompt_pilot}

\begin{table}[H]
\centering
\caption{Prompt-design pilot AUROC (550 balanced pairs, Era~1).}
\label{tab:llm_pilot}
\small
\begin{tabular}{llc}
\toprule
\textbf{Model} & \textbf{Prompt Variant} & \textbf{AUROC} \\
\midrule
Qwen-72B   & v2 (metadata only)      & \textbf{0.547} \\
Qwen-72B   & v2 + network statistics  & 0.486 \\
GPT-4o     & v2 + network statistics  & 0.467 \\
\bottomrule
\end{tabular}
\end{table}

Adding pre-computed network statistics degraded performance from 0.547 to 0.486 (Qwen-72B) and 0.467 (GPT-4o). This anchoring effect motivated the metadata-only design. Frontier models offered no improvement over open-source alternatives, suggesting prompt architecture matters more than model scale.

\section{OpenAlex API Temporal Metadata Feasibility}
\label{app:temporal_metadata}

\begin{table}[H]
\centering
\caption{OpenAlex API temporal metadata feasibility.}
\label{tab:api_feasibility}
\small
\begin{tabular}{llll}
\toprule
\textbf{Field} & \textbf{Era-Restricted?} & \textbf{Method} & \textbf{Effort} \\
\midrule
\texttt{display\_name}       & Yes (stable) & No action needed & None \\
\texttt{works\_count}        & Yes (direct) & \texttt{counts\_by\_year} & 1 call/author \\
\texttt{cited\_by\_count}    & Yes (direct) & \texttt{counts\_by\_year} & 1 call/author \\
\texttt{institution}         & Yes (direct) & \texttt{affiliations} & 1 call/author \\
\texttt{concepts/topics}     & \textbf{No (indirect)} & Per-work aggregation & Millions of calls \\
\bottomrule
\end{tabular}
\end{table}

The \texttt{counts\_by\_year} field covers the full career (verified: Bengio 1988--2026; Hinton 1976--2026). However, \textbf{research concepts are only available as cumulative career-level attributes}. Reconstructing era-restricted concepts requires per-work aggregation across millions of API calls, and produces method-dependent profiles: five aggregation methods applied to Bengio's 48 works (2004--07) yield top-5 concept profiles with 0/5 to 4/5 overlap with the career-level profile (Table~\ref{tab:concept_aggregation}).

\begin{table}[H]
\centering
\caption{Era-restricted concept profiles for Yoshua Bengio (2004--2007, 48 works) under five aggregation methods.}
\label{tab:concept_aggregation}
\small
\begin{tabular}{lll}
\toprule
\textbf{Method} & \textbf{Top-5 Concepts (2004--07)} & \textbf{Overlap} \\
\midrule
Career-level & CS, AI, ML, Math, Neural Networks & (reference) \\
\hdashline
Mean score     & Pixel, Convexity, Variation, Isomap, SGD       & 0/5 \\
Max score      & RL, Isomap, Pixel, Dim.\ Reduction, Curse of Dim. & 0/5 \\
Frequency      & CS, AI, Math, ML, Kernel                       & 4/5 \\
Citation-wtd   & CS, AI, Training, ML, Layer                    & 3/5 \\
Top-5 voting   & CS, AI, Math, Curse of Dim., ML                & 4/5 \\
\bottomrule
\end{tabular}
\end{table}

\section{Additional Results}
\label{app:additional_results}

\begin{figure}[H]
    \centering
    \includegraphics[width=0.5\textwidth]{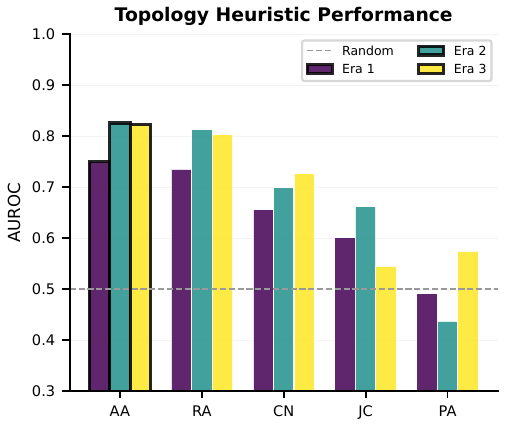}
    \caption{Topology heuristic AUROC across three eras. AA and RA consistently outperform; PA hovers near random.}
    \label{fig:heuristic_auroc}
\end{figure}

\begin{table}[H]
\centering
\caption{node2vec vs.\ topology heuristic AUROC across three eras (50,000 stratified pairs per era).}
\label{tab:node2vec_auroc}
\small
\begin{tabular}{llcccc}
\toprule
\textbf{Era} & \textbf{Best N2V Op.} & \textbf{N2V AUROC} & \textbf{Best Topo} & \textbf{Topo AUROC} & \textbf{$\Delta$} \\
\midrule
Classical AI     & L2 distance    & 0.781 & AA & 0.773 & $+$0.008 \\
DL Explosion     & Cosine         & 0.834 & AA & 0.830 & $+$0.004 \\
Democratization  & Cosine         & 0.792 & RA & \textbf{0.854} & $-$0.062 \\
\bottomrule
\end{tabular}
\end{table}

\begin{table}[H]
\centering
\caption{Continued-edge prediction AUROC (Era~1, $N=150$, 67 continued vs.\ 83 dropped).}
\label{tab:continued_edge}
\small
\begin{tabular}{lc}
\toprule
\textbf{Method} & \textbf{AUROC} \\
\midrule
Resource Allocation  & \textbf{0.706} \\
LLM (Qwen-72B)      & 0.687 \\
Adamic-Adar          & 0.684 \\
Edge Weight          & 0.659 \\
Common Neighbors     & 0.619 \\
Jaccard Coefficient  & 0.618 \\
\bottomrule
\end{tabular}
\end{table}

\begin{figure}[H]
    \centering
    \includegraphics[width=\textwidth]{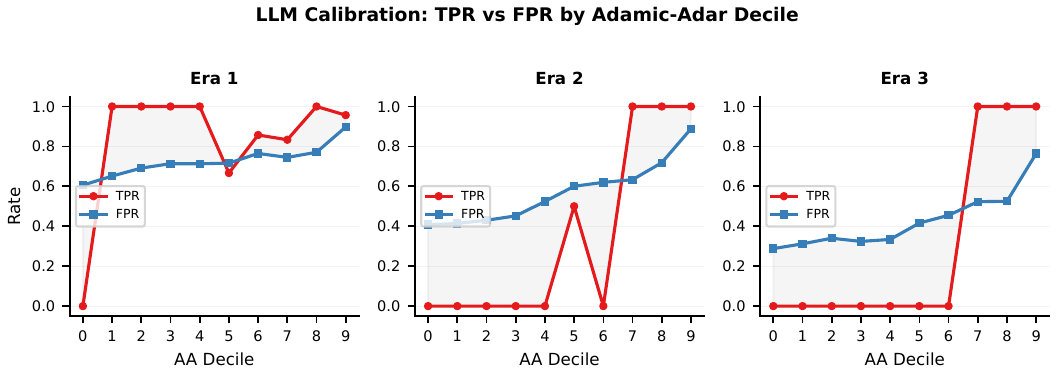}
    \caption{LLM calibration by Adamic-Adar decile across three eras. The narrow gap between TPR (red) and FPR (blue) indicates poor discrimination.}
    \label{fig:calibration_decile}
\end{figure}

\paragraph{Community structure and predictability.} Intra-community collaboration rates exceeded cross-community rates by 3.8$\times$, 3.8$\times$, and 7.0$\times$ across three eras. Continued edges were overwhelmingly intra-community, while new edges were more likely cross-community, suggesting that predicting novel, boundary-spanning collaborations is fundamentally harder than predicting tie persistence.

\section{Cold-Start Long-Tail Analysis}
\label{app:cold_start_details}

\paragraph{Degree distributions.} Authors involved in cold-start new edges have systematically lower training-graph degree than those in 2-hop-reachable new edges. The median degree gap widens from 1.2$\times$ (Era~1) to 2.6$\times$ (Era~3).

\begin{figure}[H]
    \centering
    \includegraphics[width=\textwidth]{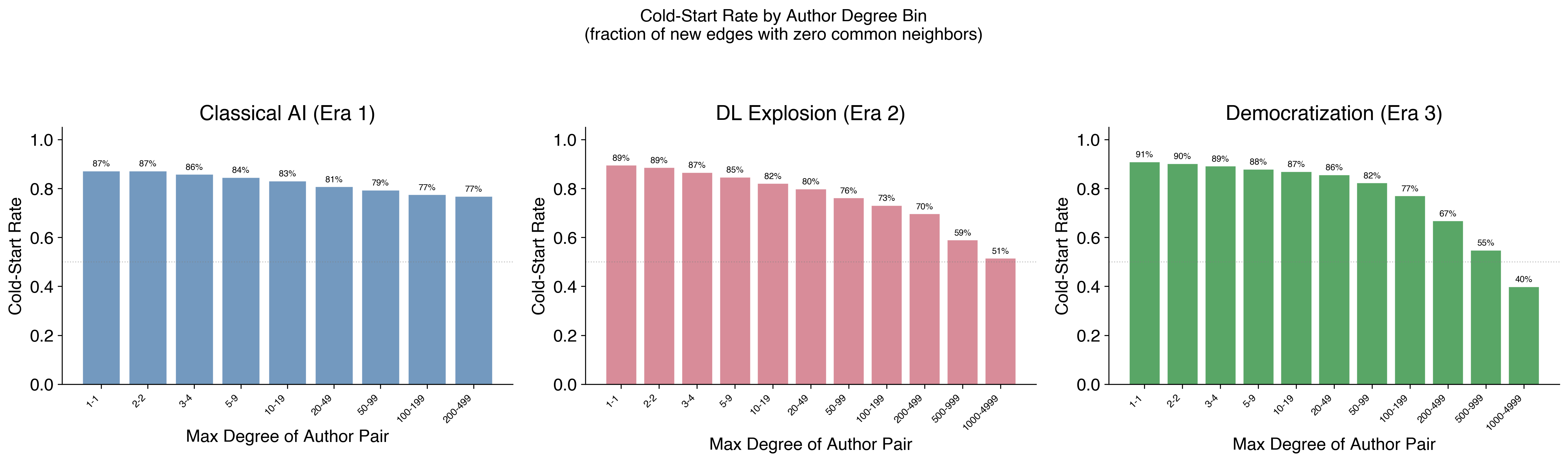}
    \caption{Cold-start rate by author degree bin. The fraction of new edges with zero common neighbors decreases monotonically with degree, from 87--91\% for degree-1 authors to 40--51\% for degree-1,000+.}
    \label{fig:cs_rate_degree}
\end{figure}

\begin{figure}[H]
    \centering
    \includegraphics[width=\textwidth]{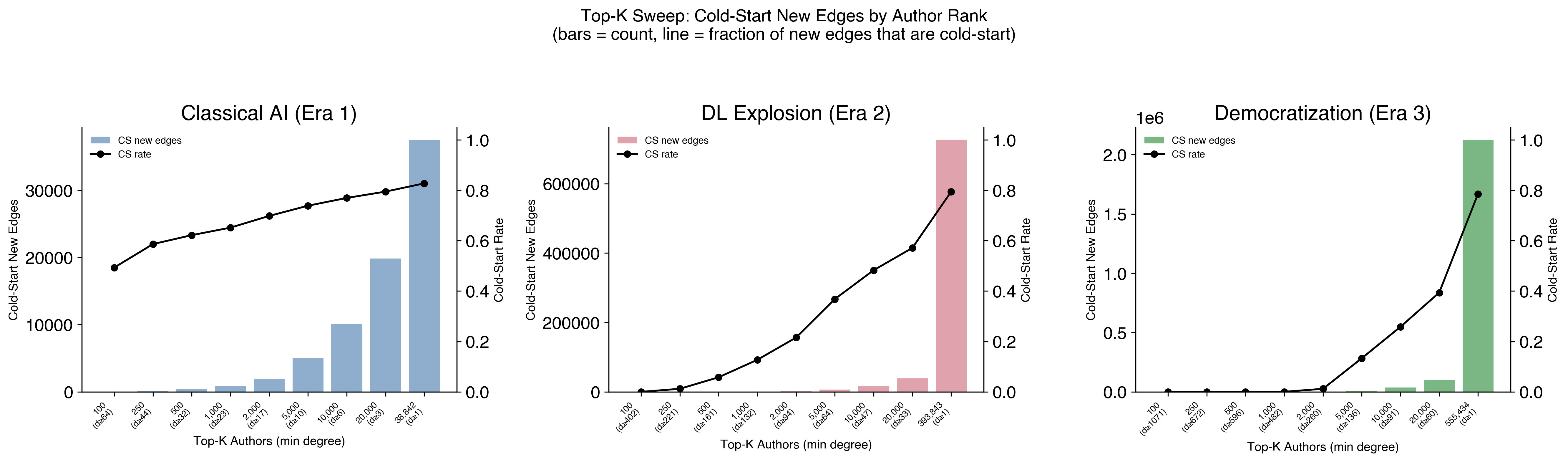}
    \caption{Top-K sweep: cold-start new edges (bars) and cold-start rate (line) by author rank. In Era~3, cold-start is absent among top-1,000 hubs but accounts for 79\% of new edges in the full network.}
    \label{fig:cs_topK}
\end{figure}

\begin{figure}[H]
    \centering
    \includegraphics[width=\textwidth]{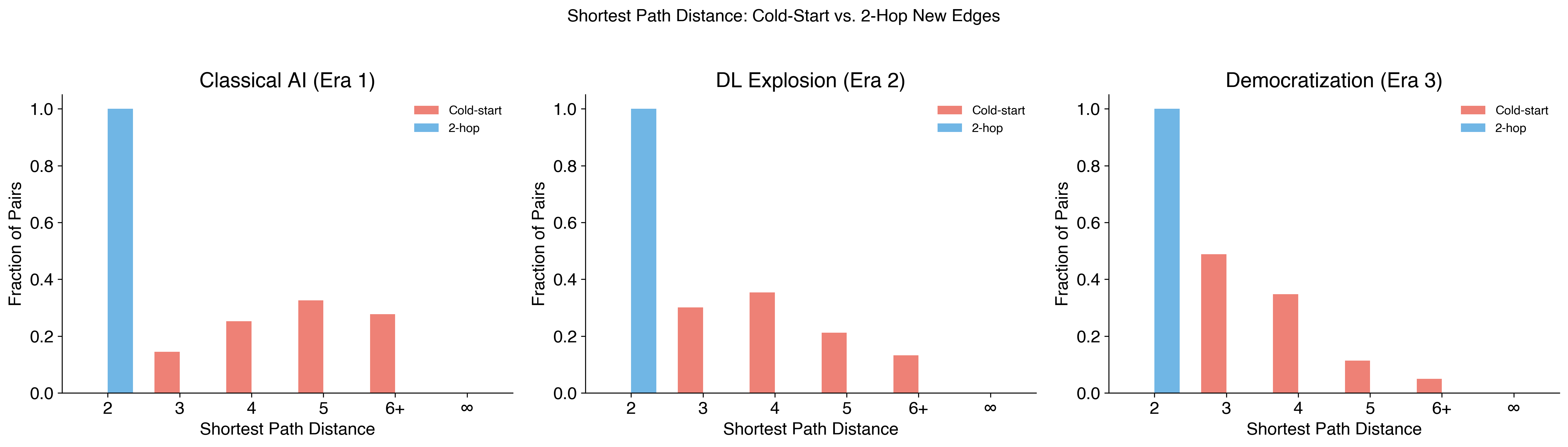}
    \caption{Shortest-path distance distribution for cold-start vs.\ 2-hop new edges. Cold-start pairs have median distance 4--5, confirming they are structurally distant.}
    \label{fig:cs_shortest_path}
\end{figure}

\begin{figure}[H]
    \centering
    \includegraphics[width=0.6\textwidth]{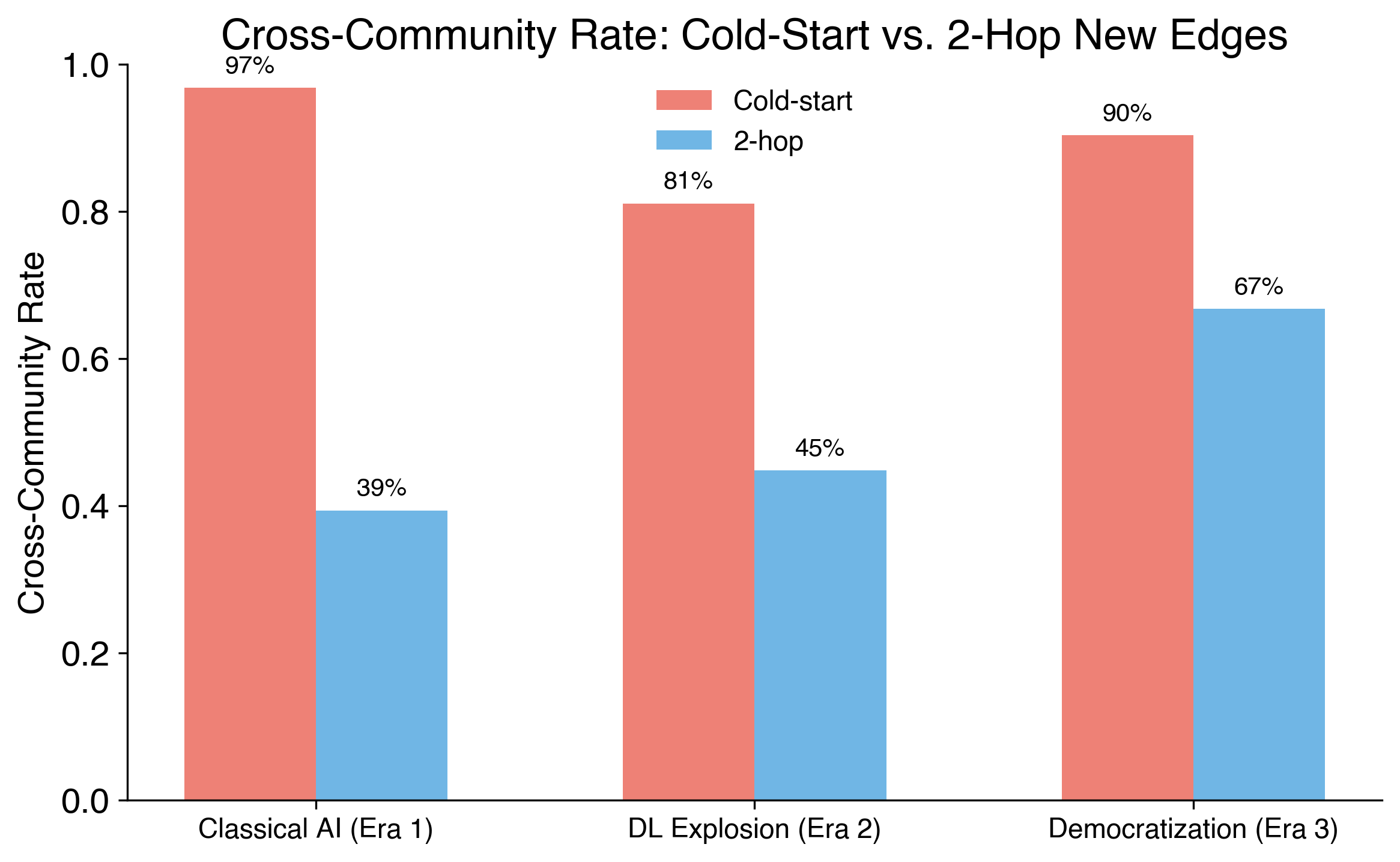}
    \caption{Cross-community rate for cold-start vs.\ 2-hop new edges. Cold-start edges are overwhelmingly cross-community (81--97\%) compared to 2-hop edges (39--67\%).}
    \label{fig:cs_cross_community}
\end{figure}

\section{Socio-Cultural Analysis Details}
\label{app:sociocultural_details}

\begin{table}[H]
\centering
\caption{Feature AUROC on Era~1 candidate pairs ($N = 5{,}000$, 47 positives). Socio-cultural features fall below random (0.5).}
\label{tab:sociocultural_auroc}
\small
\begin{tabular}{llc}
\toprule
\textbf{Category} & \textbf{Feature} & \textbf{AUROC} \\
\midrule
\multirow{2}{*}{Topology}
    & Adamic-Adar          & \textbf{0.801} \\
    & Resource Allocation  & 0.805 \\
\midrule
\multirow{2}{*}{Metadata}
    & Concept Overlap      & 0.555 \\
    & Same Institution     & 0.527 \\
\midrule
\multirow{3}{*}{Socio-cultural}
    & Same Ethnicity       & 0.475 \\
    & Same Country         & 0.448 \\
    & Same Continent       & 0.445 \\
\bottomrule
\end{tabular}
\end{table}

\begin{table}[H]
\centering
\caption{Homophily analysis on Era~1 candidate pairs. Ratios below 1.0 indicate anti-homophily.}
\label{tab:homophily}
\small
\begin{tabular}{lccc}
\toprule
\textbf{Feature} & \textbf{Collab.\ Same Rate} & \textbf{Non-collab.\ Same Rate} & \textbf{Homophily Ratio} \\
\midrule
Ethnicity  & 83.0\% & 88.0\% & 0.94 \\
Country    & 40.4\% & 50.8\% & 0.80 \\
Continent  & 46.8\% & 57.8\% & 0.81 \\
\bottomrule
\end{tabular}
\end{table}

\begin{figure}[H]
    \centering
    \includegraphics[width=\textwidth]{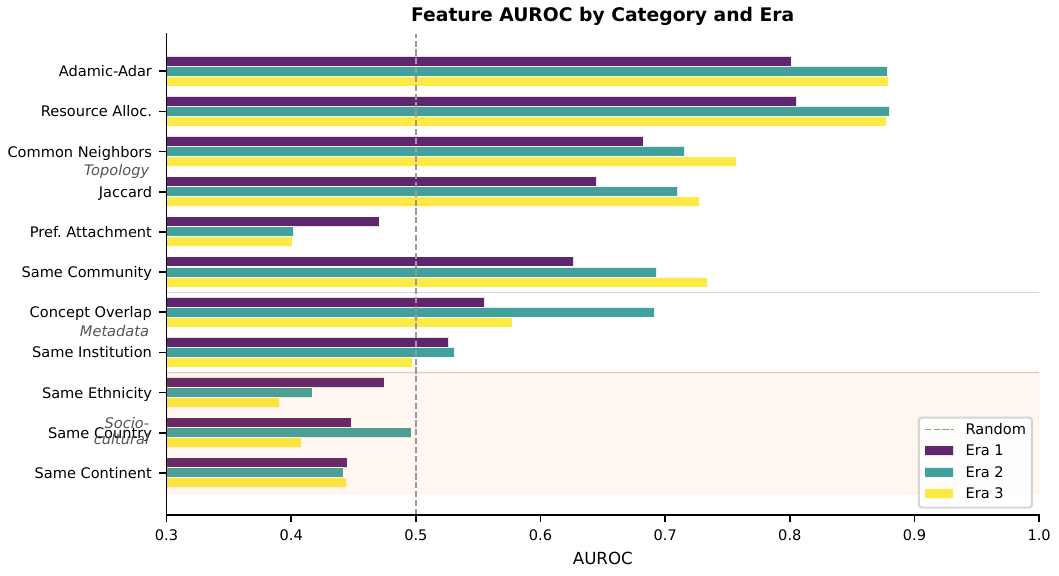}
    \caption{Feature AUROC by category and era for all 11 features. All socio-cultural features fall below the random baseline.}
    \label{fig:feature_auroc}
\end{figure}

\begin{figure}[H]
    \centering
    \includegraphics[width=\textwidth]{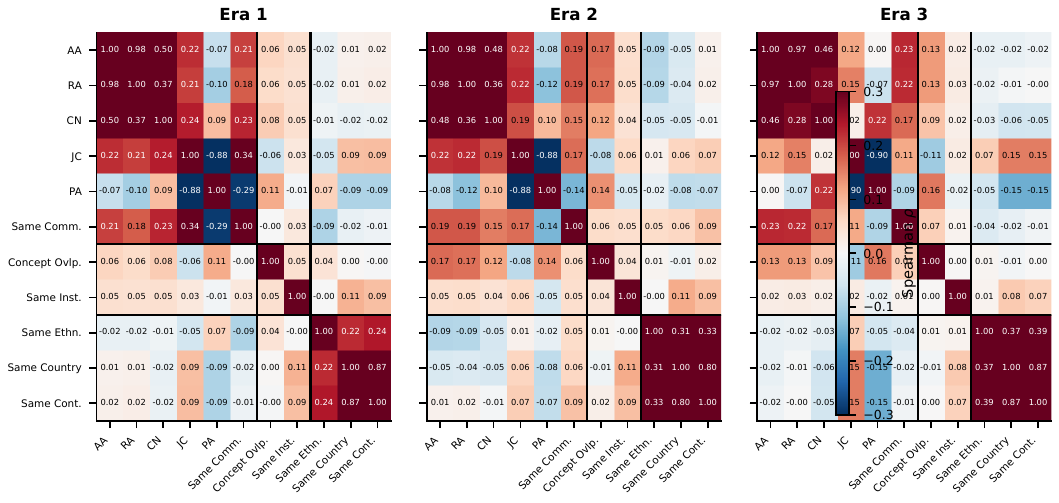}
    \caption{Spearman correlation matrix for all 11 features across three eras. Cross-category correlations are negligible ($|r| < 0.1$).}
    \label{fig:correlation_heatmap}
\end{figure}

\begin{figure}[H]
    \centering
    \includegraphics[width=\textwidth]{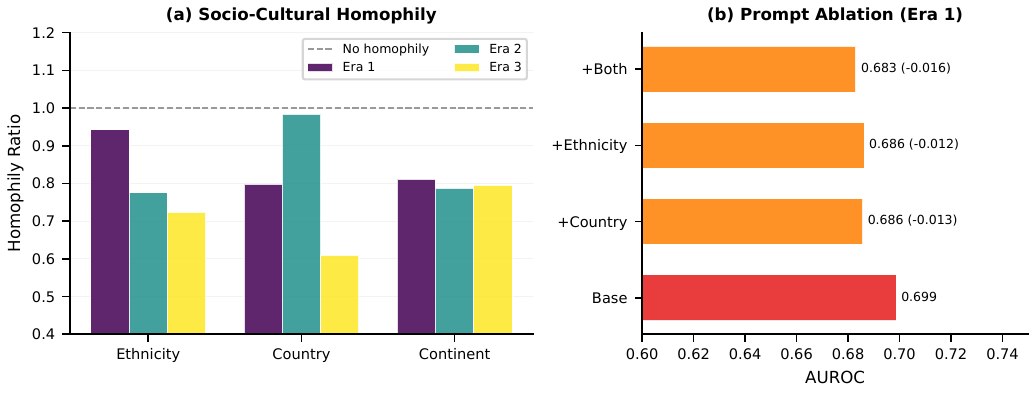}
    \caption{(a)~Homophily ratios by era: values below 1.0 indicate anti-homophily. (b)~LLM prompt ablation: adding socio-cultural context consistently degrades AUROC.}
    \label{fig:homophily_ablation}
\end{figure}

\section{Representative Prediction Examples}
\label{app:examples}

To provide qualitative grounding for the quantitative results, we present representative examples from four agreement categories between Adamic-Adar (AA) and the LLM, drawn from the evaluation samples across all three eras (68 total positives: 47 from Era~1, 14 from Era~2, 7 from Era~3). AA uses the per-era median-positive threshold; the LLM uses 0.5.

\paragraph{Case 1: Both methods succeed (35/68 positives).} These pairs exhibit strong structural \emph{and} metadata signals. For example, Xiaofeng Liao (Chongqing University of Posts and Telecommunications) and Degang Yang (Tongji University) share 10 common neighbors (AA\,=\,4.04), high concept overlap (0.67), belong to the same sub-community, and the LLM assigns probability 0.85. The convergence of structural proximity and topical similarity makes these pairs ``easy'' for both methods: median AA\,=\,0.98, median concept overlap\,=\,0.43, and 83\% are intra-community.

\paragraph{Case 2: LLM catches, AA misses (27/68 positives).} These are the pairs that demonstrate the LLM's unique value. They have low structural signal (median AA\,=\,0.32, median CN\,=\,1) but the LLM still predicts collaboration (median prob\,=\,0.65). For instance, Ivor W.\ Tsang and Changshui Zhang share perfect concept overlap (1.0) but only 1 common neighbor (AA\,=\,0.32) and are in different sub-communities---the LLM recognizes their shared machine learning focus despite minimal structural proximity. Similarly, Nitin S.\ Baliga and Ben Heavner share the same institution (University of Washington) with perfect concept overlap, but only 1 common neighbor. Only 37\% of these pairs are intra-community, compared to 83\% for ``both catch'' pairs, confirming that the LLM provides value precisely for cross-community collaborations where topology is weakest.

\paragraph{Case 3: AA catches, LLM misses (1/68 positives).} Only a single positive pair falls in this category: Guan Zhu (University of Nebraska--Lincoln) and Longin Jan Latecki (Queens College, CUNY), with AA\,=\,1.07 and 2 common neighbors but zero concept overlap and LLM probability 0.25. The LLM rejected this pair because their metadata profiles show no topical alignment, yet the structural signal (2 shared collaborators) correctly predicted the collaboration. This single case illustrates the LLM's reliance on metadata: when concept overlap is absent, the LLM cannot compensate with structural reasoning.

\paragraph{Case 4: Both methods fail (5/68 positives).} The hardest pairs have weak signals across all dimensions: median AA\,=\,0.30, median concept overlap\,=\,0.00, median LLM probability\,=\,0.35, only 1 common neighbor, and 80\% are cross-community. For example, Xing-Xiang He (Central South University) and Yue Lu (Qingdao University) share no research concepts, are in different sub-communities, and have only 1 common neighbor---neither topology nor metadata provides a basis for prediction. These pairs likely represent opportunistic or serendipitous collaborations that are inherently difficult to forecast from either information channel.

\paragraph{Implications.} The 27 ``LLM only'' pairs (40\% of all positives) directly validate the complementarity thesis: these collaborations are invisible to topology heuristics but detectable from metadata. The 5 ``both miss'' pairs (7\%) represent a floor of unpredictable collaborations. The near-absence of ``AA only'' pairs (1/68) suggests that when structural proximity is strong enough to predict collaboration, the LLM almost always agrees---the LLM's false negatives are driven by missing concept overlap, not by contradicting topology.

\section{Community Structure Visualization}
\label{app:community_vis}

For each era, we subsample up to 1,500 nodes from all Louvain communities (preserving size proportions, biased toward high-degree nodes). A two-stage layout positions communities via spring-layout on a meta-graph, then lays out each community's subgraph using Graphviz \texttt{neato} (stress minimization).

\begin{figure}[H]
    \centering
    \includegraphics[width=\textwidth]{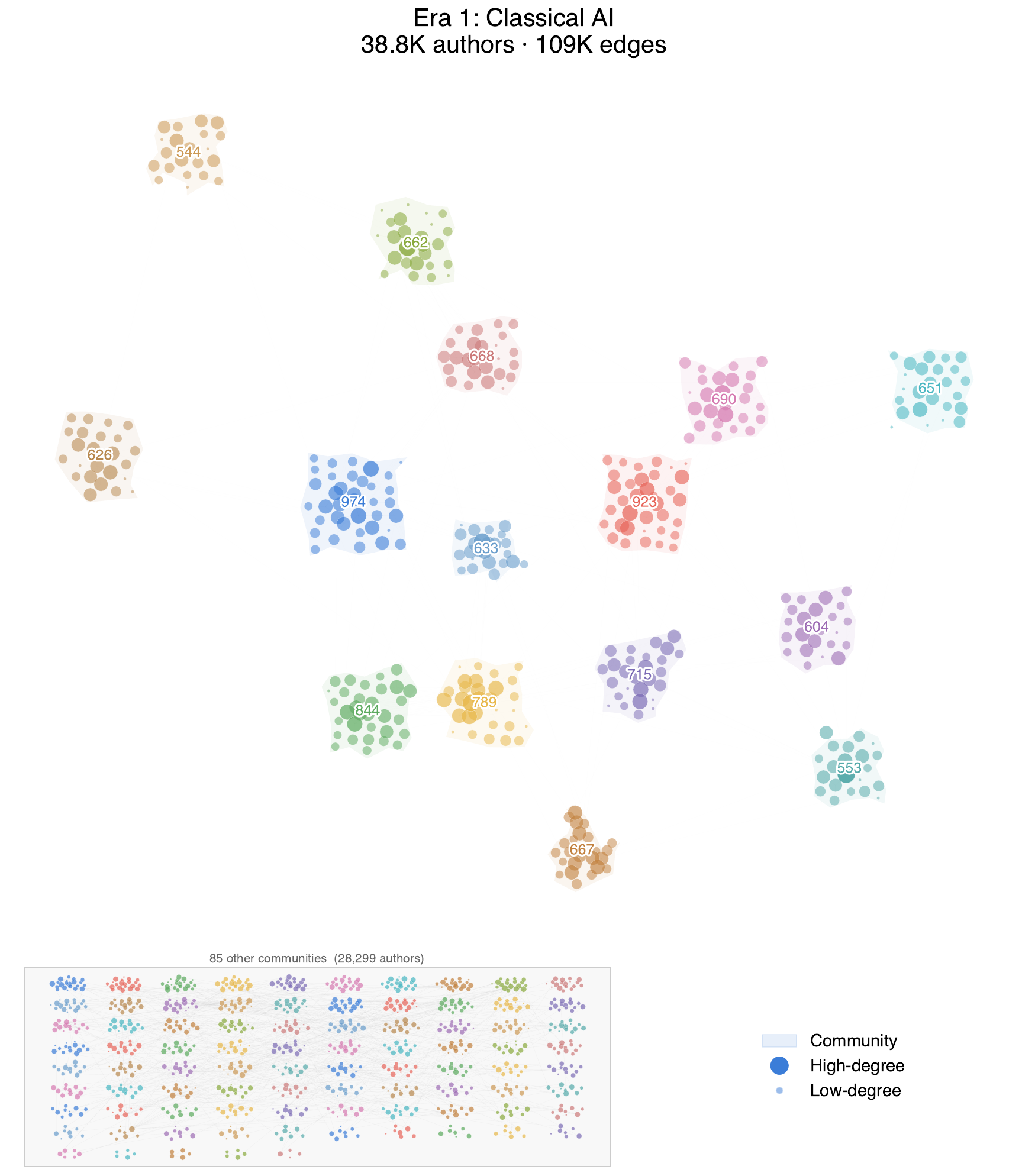}
    \caption{Community structure of the Era~1 (Classical AI) training graph (38.8K authors, 109K edges; 1,500 subsampled).}
    \label{fig:community_era1}
\end{figure}

\begin{figure}[H]
    \centering
    \includegraphics[width=\textwidth]{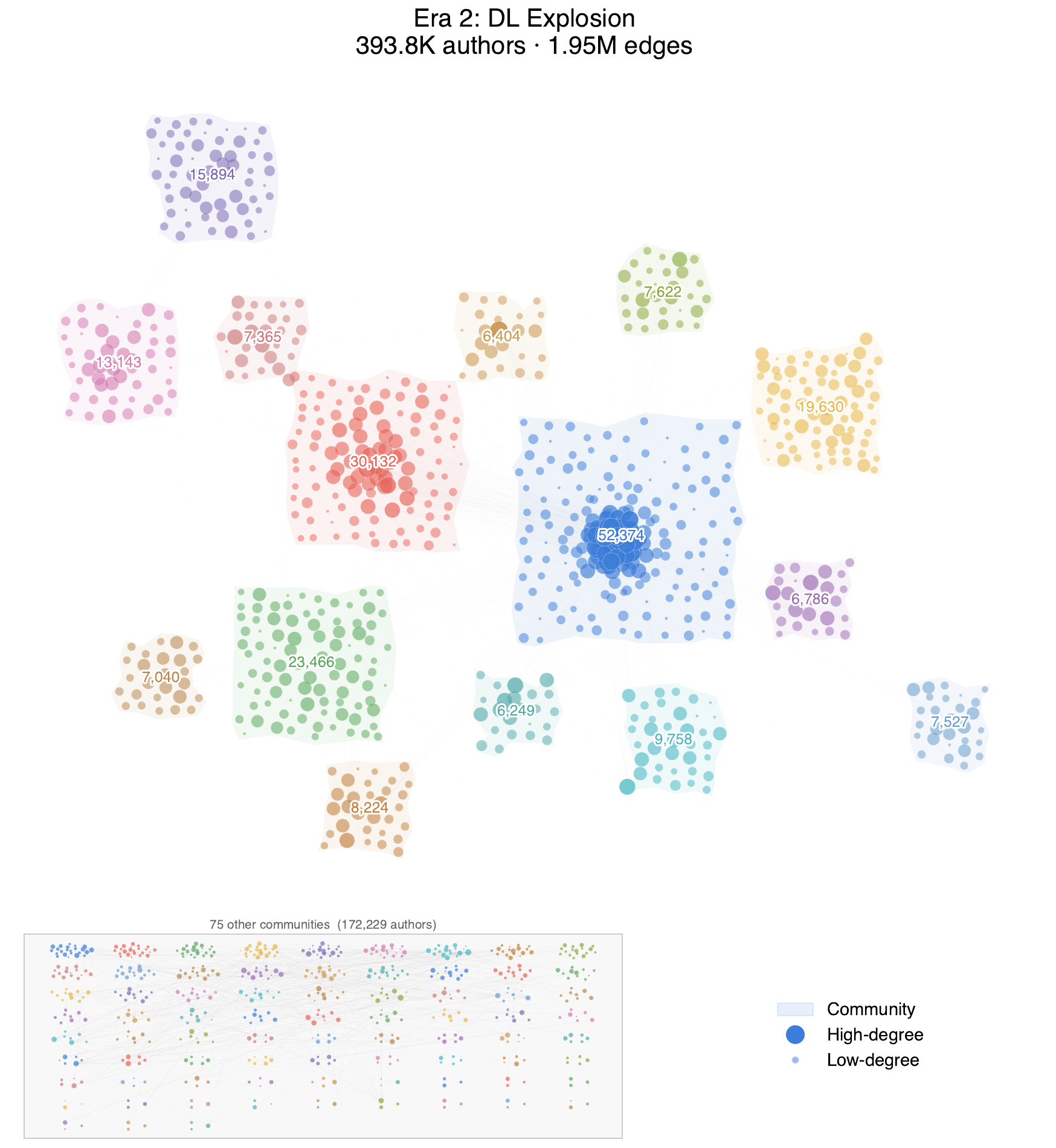}
    \caption{Community structure of the Era~2 (DL Explosion) training graph (393.8K authors, 1.95M edges; 1,500 subsampled).}
    \label{fig:community_era2}
\end{figure}

\begin{figure}[H]
    \centering
    \includegraphics[width=\textwidth]{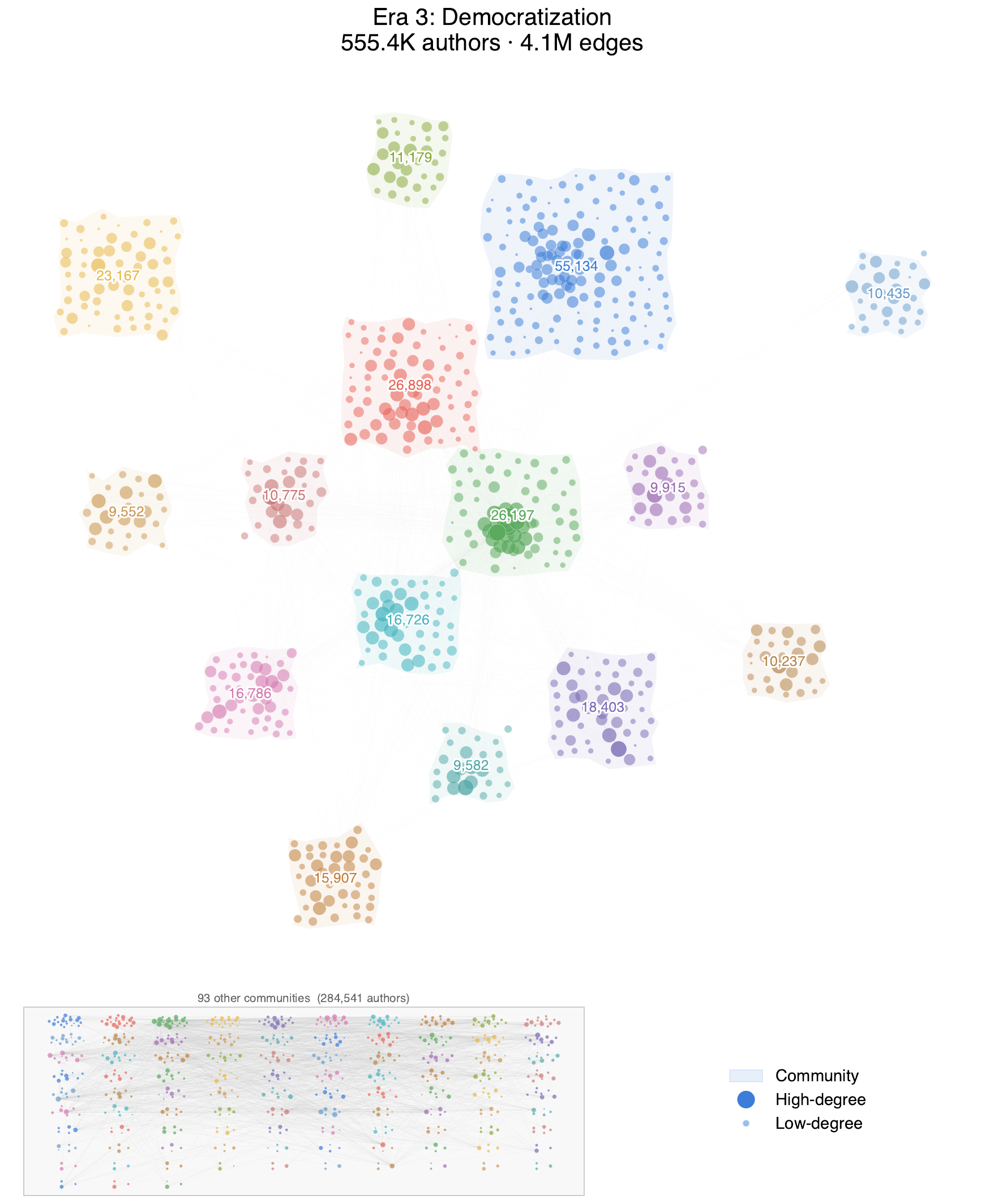}
    \caption{Community structure of the Era~3 (Democratization) training graph (555.4K authors, 4.1M edges; 1,500 subsampled).}
    \label{fig:community_era3}
\end{figure}

\end{document}